\documentclass[preprint,aps]{revtex4}
\usepackage{graphicx,epstopdf}
\begin{document}

\title[Investigating the static dipole polarizability of confined noble gas atoms]{Investigating the static dipole polarizability of noble gas atoms confined in impenetrable spheres and shells}

\author{J. A. Ludlow$\footnote{Present address: AquaQ Analytics, Suite 5, Sturgeon Building, 9-15 Queen Street, Belfast, BT1 6EA}$ and Teck-Ghee Lee}
\affiliation{Department of Physics, Auburn University, Auburn, AL 36849, USA}

\begin{abstract}
The static dipole polarizability of noble gas atoms confined by impenetrable spheres and spherical shells is studied using the B-spline random phase with exchange approximation.  The general trend in dipole polarizabilities across the noble gas sequence shows a decrease in the dipole polarizability as the volume of the confining impenetrable sphere is reduced and a large increase in the dipole polarizability for confinement by impenetrable spherical shells as the inner shell radius is increased. 
\end{abstract}

\pacs{31.15.ap, 32.10.Dk, 37.30.+i }

\maketitle

\newpage

\section{Introduction}
The study of atoms under confinement is a subject of long standing interest, for a selection of reviews see \cite{confinerev1}-\cite{confinerev6}. For many unconfined atoms, their polarizabilities are known to a high degree of accuracy, see \cite{polarrev1} and \cite{polarrev2}. An interesting question arises as to the effect of confinement on the polarizability of an atom.

The majority of studies investigating the polarizabilities of atoms under confinement have been carried out for hydrogen, see \cite{early1}-\cite{early3} for some early work and \cite{h1}-\cite{h4} for more recent calculations. For helium, an early calculation \cite{he0} has been followed by density functional \cite{he3}, configuration interaction \cite{he1}, coupled cluster \cite{he2} and variational \cite{he4} calculations with confining potentials studied including cylindrical \cite{hcylinder} and spheroidal \cite{hespheroid}. For heavier systems, one-electron model potentials have been used to study the static dipole polarizability of alkali metals under hard wall confinement \cite{alkali1} and debye screening \cite{db1}-\cite{db3}. The static dipole polarizability and hyperpolarizability of beryllium was studied under harmonic, rectangular and gaussian confining potentials using a basis of explicitly correlated gaussians \cite{be}.
For the polarizabilities of noble gas atoms under hard wall confinement, an early calculation studied the confined argon atom in a Thomas-Fermi model \cite{ar0}. Hartree-Fock calculations \cite{hartree3}-\cite{hartree5} and a time-dependent density-functional theory \cite{hartree6} calculation have since been performed along the noble gas sequence. The polarizabilities of confined noble gases are of interest in interpreting the density dependence of atomic polarizabilities \cite{nobleexp} and studies of noble gas endofullerenes \cite{endo1}-\cite{endo3}.

Here, non-relativistic random phase approximation with exchange (RPAE) calculations of the static dipole polarizability are performed for noble gas atoms under hard-wall confinement. The RPAE method \cite{amusia73}-\cite{amusia90} has been shown to give dipole polarizabilities accurate to within a few percent for the noble gases when compared to more sophisticated calculations \cite{noblecc}, with relativistic effects having a small effect on polarizabilities for the noble gases \cite{rrpa}. 

The rest of the paper is structured as follows: Section II details the RPAE approximation to the calculation of the static dipole polarizability and its numerical implemention using B-splines. In section III A, results are presented for the static dipole polarizabilities of hydrogen and noble gas atoms confined by an impenetrable sphere. Section III B investigates the case of a spherical shell potential. Finally, Section IV concludes with a summary. Unless otherwise stated, atomic units are used throughout.

\section{Theory}

\subsection{Calculation of the dipole polarizability}
For a hydrogen atom in the ground state, the static dipole polarizability can be calculated via the relation,
\begin{equation}
\label{polhy}
\alpha=2\sum_{n}\frac{|\langle np |z| 1s \rangle |^2}{\varepsilon_{np}-\varepsilon_{1s}}
\end{equation}

For noble gas atoms, electron correlation  beyond the Hartree-Fock approximation plays an important role and must be accounted for. The dipole polarizability for noble gas atoms will be calculated via the RPAE approximation  \cite{amusia73}-\cite{amusia90}. In the following equations for the RPAE vertex matrix element, the summation is over hole states, or states below the Fermi level ($\leq F$), and excited electron states above the Fermi level ( $>F$). 

For  ($\nu > F$) and ($\mu \leq F$),
\begin{equation}
\label{rpaeq1}
\langle \nu |A(\omega )| \mu \rangle = \langle \nu |d| \mu \rangle +\left (\sum_{\stackrel{\nu^{\prime}>F}{\mu^{\prime}{\leq}F}}
- \sum_{\stackrel{\mu^{\prime}>F}{\nu^{\prime}{\leq}F}}\right ) \frac{(\langle \nu\mu^{\prime} |V| \nu^{\prime}\mu \rangle
-\langle \mu^{\prime}\nu |V| \nu^{\prime}\mu \rangle)}
{\omega-\varepsilon_{\nu^{\prime}}+\varepsilon_{\mu^{\prime}}+i\delta (1-2n_{\nu^{\prime}})}
\langle \nu^{\prime} |A(\omega )| \mu^{\prime} \rangle , 
\end{equation}
and for ($\nu \leq F$), ($\mu >F$),
\begin{equation}
\label{rpaeq2}
\langle \nu |A(\omega )| \mu \rangle = \langle \nu |d| \mu \rangle +\left (\sum_{\stackrel{\nu^{\prime}>F}{\mu^{\prime}{\leq}F}}
- \sum_{\stackrel{\mu^{\prime}>F}{\nu^{\prime}{\leq}F}}\right ) \frac{(\langle \nu\mu^{\prime} |V| \nu^{\prime}\mu \rangle
-\langle \mu^{\prime}\nu |V| \nu^{\prime}\mu \rangle)}
{\omega-\varepsilon_{\nu^{\prime}}+\varepsilon_{\mu^{\prime}}+i\delta (1-2n_{\nu^{\prime}})}
\langle \nu^{\prime} |A(\omega )| \mu^{\prime} \rangle ,
\end{equation}
where $i\delta$ is an infinitesimal imaginary positive, and
\begin{eqnarray}
n_{\nu^{\prime}}=\left\{\begin{array}{lll}
0 & \mbox{for}& \nu^{\prime} > F\\
1 & \mbox{for}& \nu^{\prime} \leq F \,\,\,\, .
\end{array}\right.
\end{eqnarray}
Numerically, equations (\ref{rpaeq1}) and (\ref{rpaeq2}) are a set of linear algebraic equations, which can be written in block matrix form,
\begin{equation}
\left(\begin{array}{l}
x\\
y
\end{array}\right) = \left(\begin{array}{l}
d\\
d
\end{array}\right) + \left(\begin{array}{ll}
V_1&V_2\\
V_2&V_1
\end{array}\right)\left(\begin{array}{ll}
\chi_1&0\\
0&\chi_2
\end{array}\right)\left(\begin{array}{l}
x\\
y
\end{array}\right) ,
\end{equation}
where $\chi_1$ and $\chi_2$ are energy denominators, $x$ and $y$ represent $\langle \nu |A(\omega )| \mu \rangle $ for 
$\nu > F$, $\mu\leq F$ and $\nu\leq F$, $\mu > F$ respectively, $d$ are the Hartree-Fock dipole matrix elements and $V_1$, $V_2$ are Coulomb matrices.

Once this equation has been solved for the RPA vertex, the dipole polarizability can be calculated via the relation:
\begin{equation}
\label{rpaeq3}
\alpha(\omega)=\sum_{\stackrel{\nu>F}{\mu{\leq}F}}\frac{\langle \mu |d| \nu \rangle \langle \nu |A(\omega )| \mu \rangle }{\omega-\varepsilon_{\nu}+\varepsilon_{\mu}+i\delta}
+\sum_{\stackrel{\nu{\leq}F}{\mu>F}}\frac{\langle \mu |d| \nu \rangle\langle \nu |A(\omega )| \mu \rangle }{-\omega+\varepsilon_{\nu}-\varepsilon_{\mu}+i\delta} \,\,\, .
\end{equation}
In this work the static dipole polarizability, $\omega=0$, is calculated.

\subsection{Numerical Implementation}
The Hartree-Fock equations will be solved for noble gas atoms confined by a potential. Two forms of the confining potential will be compared here. First, an impenetrable spherical potential of the form,
\begin{equation}\label{sphpot1}
V_{sph}(r) = \left\{ \begin{array}{ll} 
              0         & {\rm for}\,\,  r < r_c \\
               \infty & {\rm for}\,\, r\geq  r_c
             \end{array} \right .
\end{equation}
where $r_c$ is the radius of the confining potential. Secondly, an impenetrable spherical shell potential of the form.
\begin{equation}\label{sphpot2}
V_{shell}(r) = \left\{ \begin{array}{ll} 
               \infty & {\rm for}\,\, 0 \leq r \leq r_{\rm in} \\
             0         & {\rm for}\,\,  r_{\rm in} < r < r_{\rm out} \\
               \infty & {\rm for}\,\, r\geq  r_{\rm out}
             \end{array} \right .
\end{equation}
where $r_{\rm in}$ is the inner radius and $r_{\rm out}$ is the outer radius of the spherical shell potential. As the radius of the confining potential may lie within the radius of the unconfined atom, it is necessary to incorporate potentials (\ref{sphpot1}) and (\ref{sphpot2}) into the numerical solution of the Hartree-Fock equations \cite{hfcode1},\cite{hfcode2}. For all noble gas atoms, the Hartree-Fock equations were solved using a cut off radius of 30 a.u. and 6000 integration points. Note that for heavier atoms, convergence of the Hartree-Fock equations for small radii of the confining potential is more problematic. Results will be presented in this work down to the minimum radius of the confining potential for which convergence was obtained.

As polarizabilities of atoms can involve substantial contributions from continuum states, an efficient means of spanning the single particle energy continuum is needed. Here, B-spline basis sets will be used \cite{bspline, deboor}. B-splines are piecewise polynomials defined on a particular knot sequence. For confinement by an impenetrable spherical potential, an exponential knot sequence is used of the form,
\begin{equation}
t_1=t_2={\cdots}=t_k=0 ,
\end{equation}
\begin{equation}
t_{n+1}=t_{n+2}={\cdots}=t_{n+k}=r_c .
\end{equation}
For $t_{k+1}{\rightarrow}t_n$,
\begin{equation}
t_{i}=r_0\left (\exp({\beta}(i-k))-1\right ) ,
\end{equation}
\begin{equation}
{\beta}=\frac{\ln(\frac{r_c}{r_0}+1)}{(n+1-k)} ,
\end{equation}
where $ r_0$ is a parameter which affects the smallest spacing of the knot sequence.
For confinement by an impenetrable shell potential, an exponential knot sequence is used of the form,
\begin{equation}
t_1=t_2={\cdots}=t_k=r_{\rm in} ,
\end{equation}
\begin{equation}
t_{n+1}=t_{n+2}={\cdots}=t_{n+k}=r_{\rm out} .
\end{equation}
For $t_{k+1}{\rightarrow}t_n$,
\begin{equation}
t_{i}=r_{\rm in}+r_0\left (\exp({\beta}((i-k))-1\right ) ,
\end{equation}
\begin{equation}
{\beta}=\frac{\ln(\frac{(r_{\rm out}-r_{\rm in})}{r_0}+1)}{(n+1-k)} .
\end{equation}
A value of $r_0=0.001$, with $n=60$ splines of order $k=9$ was used in this work.

By expanding the radial wavefunctions in terms of B-splines,
\begin{equation}\label{radwf}
P_l(r)=\sum_ic_i^{(l)}B_i(r) ,
\end{equation}
and substituting into the atomic hamiltonian for hydrogen or the Hartree-Fock equation for noble gas atoms,
\begin{equation}
H^{(l)}P_l(r)={\epsilon}P_l(r) ,
\end{equation}
a generalised eigenvalue problem is obtained  for angular momentum $l$ that can be solved to obtain a complete set of states suitable for the calculation of the atomic polarizabilities via equations (\ref{polhy})-(\ref{rpaeq3}). The present calculations were carried out up to a maximum orbital angular momentum of $l=10$.
For confinement by an impenetrable spherical sphere, the boundary conditions $P_l(0)=P_l(r_c)=0$ are implemented by removing the first and last splines, leaving n-2 electron eigenstates for each orbital angular momentum $l$.
Similarly, for confinement by an impenetrable spherical shell, the boundary conditions $P_l(r_{\rm in})=P_l(r_{\rm out})=0$ are also implemented by removing the first and last splines, leaving n-2 electron eigenstates for each orbital  angular momentum $l$.
 For more details on the numerical implementation of the RPAE approximation, see chapter 2 of \cite{ludlowphd}.

\clearpage

\section{Results}

\subsection{Static dipole polarizabilities of atoms confined by an impenetrable spherical potential}
In order to benchmark the B-spline basis used in this work, equation (\ref{polhy}) is used to calculate the static dipole polarizability of a hydrogen atom confined in an impenetrable sphere, with the results tabulated in table \ref{stpol1}. Comparison with the results of perturbation-Hylleraas \cite{h4} and perturbation-numerical \cite{h3} calculations shows excellent agreement with the B-spline results. Comparison with the lower bound Kirkwood \cite{h4} and Buckingham \cite{h4} results shows that the higher order terms included in the Buckingham approximation \cite{early4} leads to a more accurate lower bound on the polarizability than the Kirkwood approximation \cite{early5}.

\begin{table}[ht!]
\small
\begin{center}
\caption{Static dipole polarizabilities for a hydrogen atom confined in an impenetrable sphere.}
\label{stpol1} 
\vspace{12pt}
\begin{tabular}{llllll}
R   & Present results & Kirkwood \cite{h4} & Buckingham \cite{h4} &Hylleraas \cite{h4} & Perturbation \cite{h3} \\
\hline
2.0 &   0.34255811   &  0.3401420  & 0.3401565   &  0.3425581   &  0.342558   \\
4.0 &   2.37798233   &  2.2908060  & 2.3578222   &  2.3779823   &  2.377982   \\
6.0 &   4.05814049   &  3.7054757  & 4.0471005   &  4.0581405   &  4.058140   \\
8.0 &   4.45396473   &  3.9750587  & 4.4527846   &  4.4539647   &  4.453965   \\
\hline
\end{tabular}
\end{center}
\end{table}

To calculate the polarizability for multi-electron atoms via the RPAE approximation, bound states found from solving the Hartree-Fock equations are used to construct a complete set of states in a B-spline basis. Tables \ref{stpol2} and \ref{stpol3} compares the present Hatree-Fock subshell energies for helium and neon with the Roothaan-Hartree-Fock results of \cite{hartree1}. For the energy of the 1s subshell of helium there is good agreement between the calculations, while for the 1s, 2s and 2p subshells of neon, the present subshell energies are more deeply bound than those of \cite{hartree1}. To further investigate this discrepancy, Roothaan-Hartree-Fock calculations for Neon similar to those of \cite{hartree1} could be performed, examining convergence of the subshell energies with respect to both the orbital basis set size and the optimisation of the basis set exponents.
\begin{table}[ht!]
\small
\begin{center}
\caption{Hartree-Fock energies for the 1s subshell of He  confined in an impenetrable sphere.}
\label{stpol2} 
\vspace{12pt}
\begin{tabular}{lll}
R &    E$_{\rm 1s}$  (present work) &  E$_{\rm 1s}$  \cite{hartree1}         \\
\hline
2.0    &   -0.65754880      &  -0.6593                            \\ 
4.0    &   -0.91252115      &  -0.9125               \\
6.0    &   -0.91787985      &  -0.9174                  \\
\hline
\end{tabular}
\end{center}
\end{table}

\begin{table}[ht!]
\small
\begin{center}
\caption{Hartree-fock energies for the 1s, 2s and 2p subshells of Ne  confined in an impenetrable sphere.}
\label{stpol3} 
\vspace{12pt}
\begin{tabular}{lllllll}
R &  1s (present work) &  E$_{\rm 1s}$ \cite{hartree1} &   E$_{\rm 2s}$ (present work) &  E$_{\rm 2s}$ \cite{hartree1} &  E$_{\rm 2p}$ (present work) &  E$_{\rm 2p}$ \cite{hartree1}   \\
\hline
2.5    &  -32.41538078       &  -32.3831   &  -1.72747430     &  -1.7084 &  -0.62658124    &  -0.6076      \\
3.0    &  -32.63113594        & -32.5892    & -1.84794644     & -1.8239  &  -0.76329739  & -0.7386\\
3.5    &  -32.71673490        & -32.6763    & -1.89664949     & -1.8731  &  -0.81625604 & -0.7924 \\
4.0    &  -32.75109004       &  -32.7166   &  -1.91699524     & -1.8963  &  -0.83731963 & -0.8169 \\
\hline
\end{tabular}
\end{center}
\end{table}

\begin{figure}[ht!]
\begin{center}
\includegraphics[width=\textwidth,trim={0 0 -50 -125},clip=true]{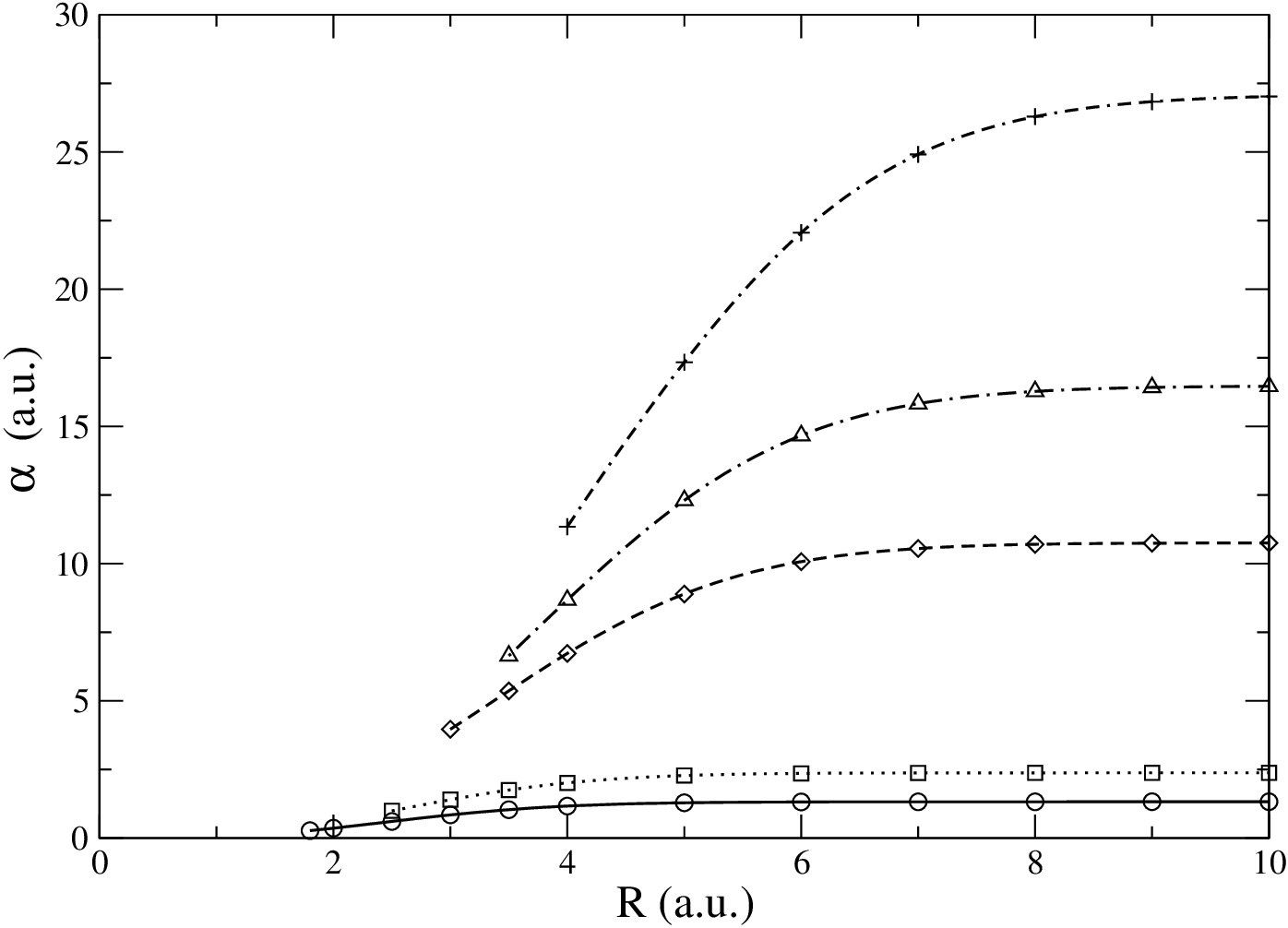}
\caption{\label{fig1} 
Static dipole polarizabilities of the noble gas sequence as a function of the confining potential radius. Solid line with circles, helium; dotted line with squares, neon; dashed line with diamonds, argon; dot-dashed line with triangles, krypton;
dot-dashed--dashed line with pluses, xenon. 
}
\end{center}
\end{figure}

Table \ref{stpol4} contains static dipole polarizabilities calculated in the RPAE approximation for noble gas atoms confined in an impenetrable sphere in comparison to unconfined polarizabilities, with the results displayed graphically in figure \ref{fig1}.
It is seen that as the confining potential radius increases, the polarizability converges smoothly to closely agree with the relativistic RPA calculations of \cite{rrpa}, demonstrating the small influence on the polarizabilities of relativistic effects. The non-relativistic RPA calculations of \cite{amusia73} are slightly lower than the present results with the exception of xenon where the results of \cite{amusia73} are higher. The differences are within the estimated 5\% uncertainty of the earlier calculations \cite{amusia73}.
As the confining potential radius decreases, the polarizability decreases monotonically. Comparison with more precise
hylleraas calculations for helium \cite{hylleraas} and relativistic coupled cluster calculations for neon, argon, krypton and xenon \cite{noblecc} shows that the non-relativisitic RPAE method is able to recover the static dipole polarizability to within a few percent.

\begin{table}[ht!]
\small
\begin{center}
\caption{Static dipole polarizabilities calculated in the RPAE approximation for nobles gas atoms confined in an impenetrable sphere.}
\label{stpol4} 
\vspace{12pt}
\begin{tabular}{llllll}
R   & He & Ne & Ar & Kr & Xe\\
\hline
1.8 &  0.27149963     &                &                  &     &   \\
2.0 &  0.35985693     &                &                  &     &  \\ 
2.5 &  0.60574486     &  0.99099895  &                  &     &   \\
3.0 &  0.84391160     &  1.39793917  &  3.96443519    &     &     \\
3.5 &  1.03453905     &  1.74913386  &  5.36273398    &   6.64086248   &     \\
4.0 &  1.16491541     &  2.00968621  &  6.72840349    &   8.67526518   & 11.34371382   \\
5.0 &  1.28514057     &  2.27781370  &  8.89416026    &  12.30276614   & 17.33513804   \\
6.0 &  1.31534685     &  2.35591238  & 10.07330554    &  14.67047514   & 22.05689052   \\
7.0 &  1.32113843     &  2.37303121  & 10.54979313    &  15.82775188   & 24.91468796    \\
8.0 &  1.32206986     &  2.37613351  & 10.70305446    &  16.27562844   & 26.28682580  \\
9.0 &  1.32220435     &  2.37663254  & 10.74468581    &  16.42077761   & 26.83277612   \\
10.0 & 1.32222378     &  2.37671053  & 10.75476508    &  16.46192981   & 27.02021394 \\
15.0 & 1.32223132     &  2.37673811  & 10.75782869    &  16.47631699   & 27.10088802  \\
20.0 & 1.32223265     &  2.37674263  & 10.75792504    &  16.47654554   & 27.10156072  \\
30.0 & 1.32223336     &  2.37674498  & 10.75797675    &  16.47666561   & 27.10188992 \\
\hline
RPAE \cite{amusia73} &    & 2.30   &  10.73  & 16.18  &  27.98 \\
RRPA\cite{rrpa} & 1.322      &   2.38    &          10.77     &   16.47          &            26.97 \\
RCCSDT\cite{noblecc} & & 2.697& 11.22 &16.80 & 27.06 \\
Hylleraas \cite{hylleraas} &1.38376079 &&&&\\
\hline
\end{tabular}
\end{center}
\end{table}

\begin{figure}[ht!]
\begin{center}
\includegraphics[width=\textwidth,trim={0 0 -50 -125},clip=true]{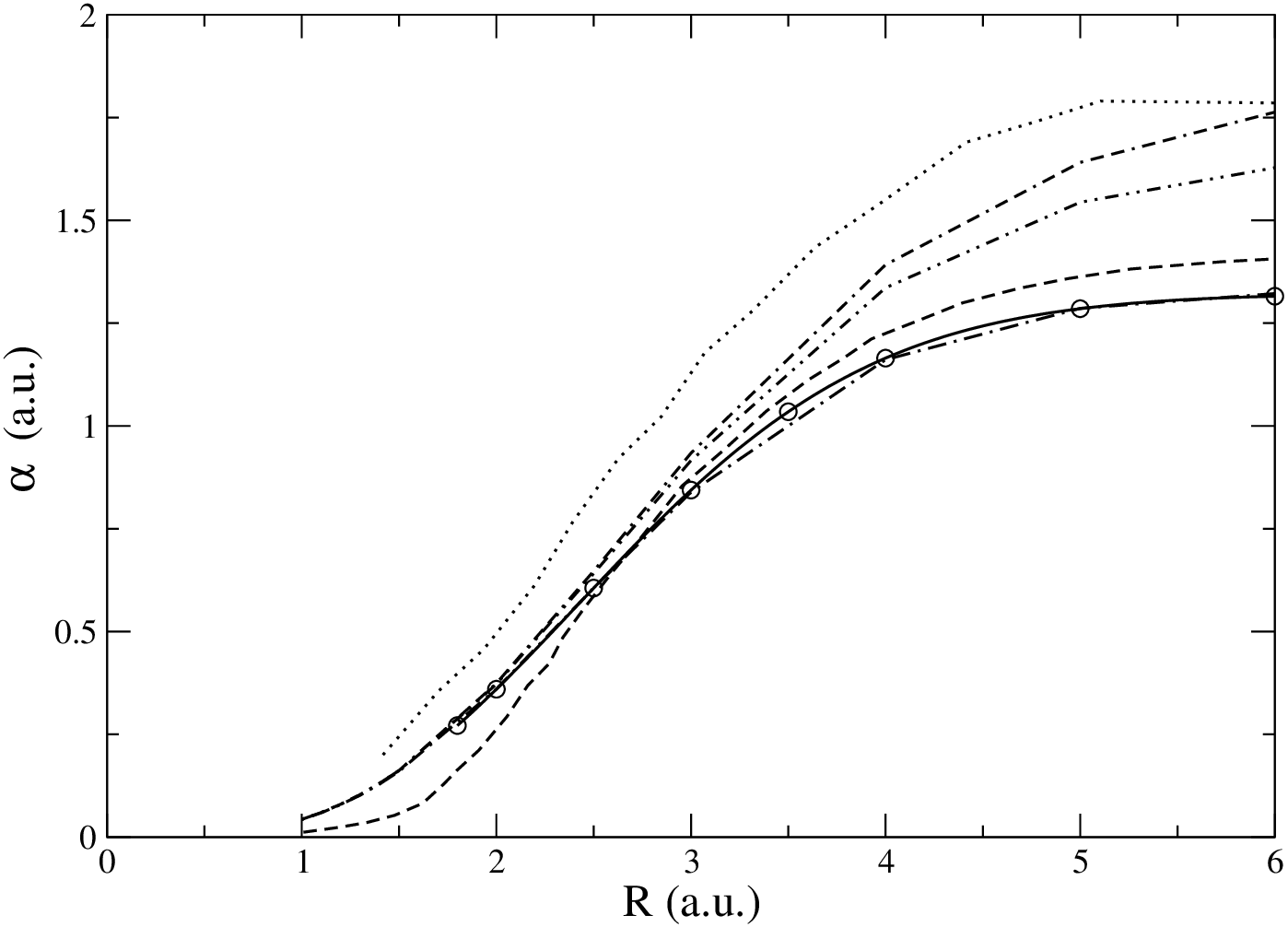}
\caption{\label{fig2} 
Static dipole polarizabilities of helium as a function of the confining potential radius. Solid line with circles, present results; dotted line, HF \cite{hartree3} (digitized using g3data \cite{g3data}); dashed line, TDDFT \cite{hartree6} (digitized using g3data \cite{g3data});
dot-dashed line, DFT (EXX) \cite{he3}; dot-dashed--dashed line, DFT (XO-LDA) \cite{he3}; dot-dot-dashed, DFT  (XC-LDA) \cite{he3}.
}
\end{center}
\end{figure}

In figure \ref{fig2}, the present RPAE polarizabilities for helium are compared to Hartree-Fock (HF) \cite{hartree3}, density functional theory (DFT) \cite{he3} and time-dependent density-functional theory (TDDFT) \cite{hartree6} calculations. The Hartree-Fock polarizabilities of \cite{hartree3} are larger than the current results and the other calculations at all box radii. The time-dependent density-functional theory polarizabilities of \cite{hartree6} are higher than the current results above a confining potential radius of 3 a.u. and fall off more strongly below a confining potential radius of 3 a.u. Excellent agreement with the present results is seen with the exact exchange, EXX, density functional results of \cite{he3}. The exchange-only local-density approximation, XO-LDA, and exchange-correlation local-density approximation, XC-LDA, polarizabilities trend higher than the present results for larger confining potential radii, but agree better for smaller radii.

\begin{figure}[ht!]
\begin{center}
\includegraphics[width=\textwidth,trim={0 0 -50 -125},clip=true]{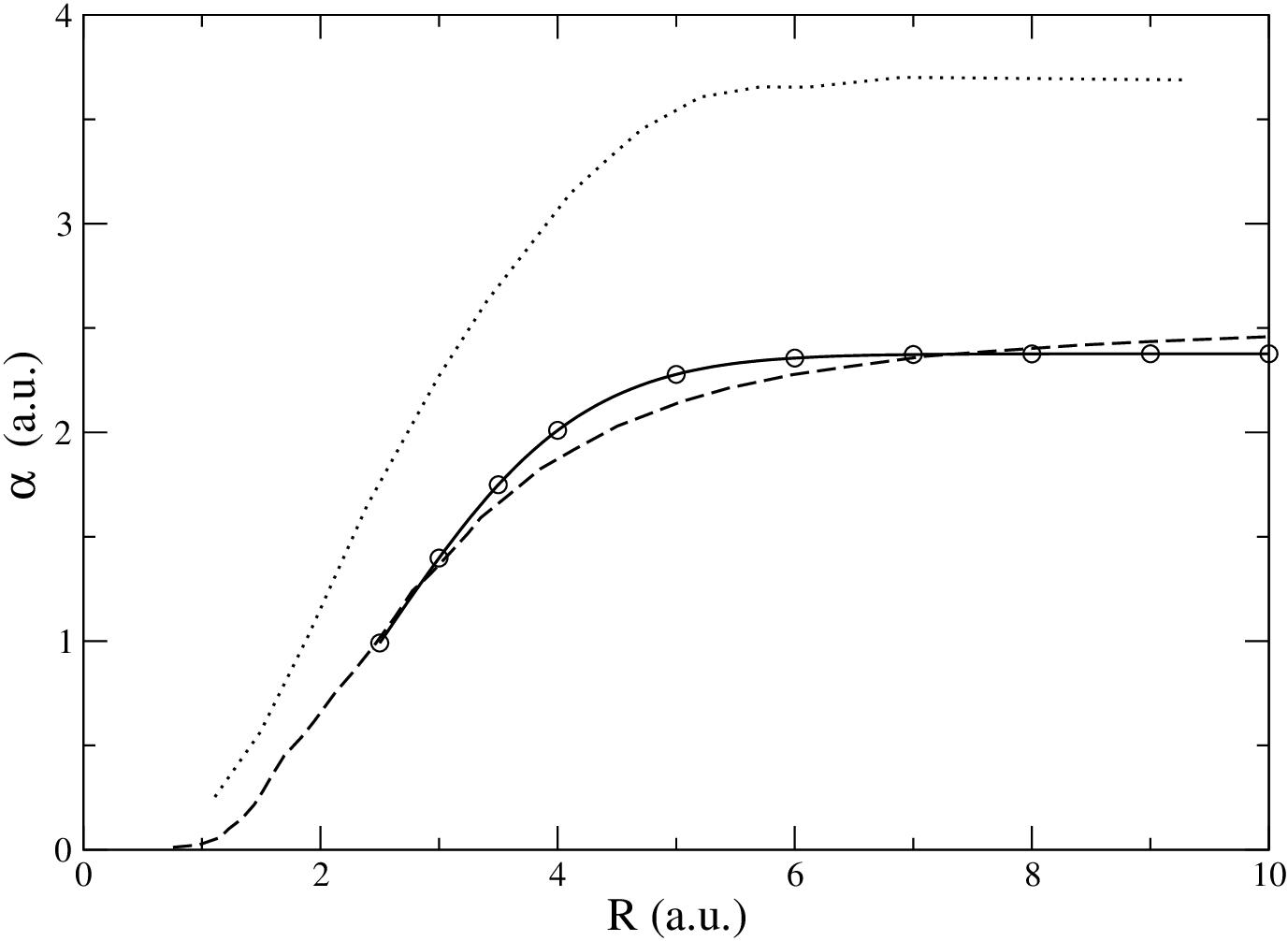}
\caption{\label{fig3} 
Static dipole polarizabilities of neon as a function of the confining potential radius. Solid line with circles, present results; dotted line, \cite{hartree3} (digitized using g3data \cite{g3data}); dashed line, \cite{hartree6}  (digitized using g3data \cite{g3data}).
}
\end{center}
\end{figure}

For Neon, figure \ref{fig3} compares the present RPAE polarizabilities to Hartree-Fock \cite{hartree3} and time-dependent density-functional theory \cite{hartree6} calculations. Similarly to the case of helium, the  Hartree-Fock polarizabilities of \cite{hartree3} are substantially higher than the current results. The  time-dependent density-functional theory polarizabiltiies of \cite{hartree6} are in good agreement with the present results, with a distinctive sigmoidal shape seen in the polarizabilities as a function of the confining potential radius.

\clearpage

\subsection{Static dipole polarizabilities of atoms confined by an impenetrable spherical shell potential}
It has been demonstrated that confinement of  noble gas atoms by an impenetrable spherical potential leads to a monotonic decrease in the polarizability as the radius, and hence volume of the confining potential is reduced.  A natural extension is to an impenetrable spherical shell potential, c.f. equation (\ref{sphpot2}). This potential may be of interest to the case of an atom confined by a negatively charged fullerene shell \cite{chbucky}.
The polarizability of hydrogen confined in an impenetrable shell potential  has been studied in \cite{hshell1}-\cite{hshell5}. Here, the confinement of helium and neon in a shell potential will be examined.

As for the case of an impenetrable spherical potential, the present calculations are first benchmarked for hydorgen. Table \ref{shellh} compares the static dipole polarizabilities calculated using equation  (\ref{polhy}) against existing calculations \cite{hshell1} for hydrogen confined in a spherical shell potential of outer radius 10 a.u. with a varying inner shell radius. There is seen to be close accord between the present results and those obtained in \cite{hshell1} using a pertubation numerical method and the Buckingham approximation  \cite{early4}. Similarly to confinement by an impenetrable spherical potential, results obtained using the Kirkwoord approximation  \cite{early5} tend to underestimate the polarizabilities. The polarizabilities increase rapidly as the inner shell radius is increased.

\begin{table}[ht!]
\small
\begin{center}
\caption{Static dipole polarizabilities for hydrogen confined by an impenetrable spherical shell.}
\label{shellh} 
\vspace{12pt}
\begin{tabular}{llllll}
$r_{\rm in}$ &  $r_{\rm out}$     & Present results & Perturbation \cite{hshell1} &   Kirkwood \cite{hshell1}  & Buckingham \cite{hshell1}      \\
\hline
2.0    &  10.0       &     485.28323894          &  474.3865   &   420.8943  &   465.6254            \\ 
4.0    &  10.0       &    1393.53470841          &  1372.0345  &   1043.0354 &   1352.7657              \\
\hline
\end{tabular}
\end{center}
\end{table}

RPAE calculations of the polarizabilities of helium and neon confined by a spherical shell potential of outer radius 10 a.u. with a varying inner shell radius are presented in table \ref{shellhene}.
The trend in the polarizabilities for helium and neon is similar to the case of hydrogen, with the polarizability for neon rising more strongly as the inner shell radius is increased than that of helium. 

\begin{table}[ht!]
\small
\begin{center}
\caption{Static dipole polarizabilities calculated in the RPAE approximation for helium and neon confined by an impenetrable spherical shell.}
\label{shellhene} 
\vspace{12pt}
\begin{tabular}{llllll}
$r_{\rm in}$ &  $r_{\rm out}$     & He                           &      Ne             \\
\hline
0.0    &  10.0       &     1.32222378             &     2.37671053                 \\
0.25   &  10.0       &    12.16238797             &    73.25472372                \\ 
0.5    &  10.0       &    31.94196072             &   164.42341795                 \\
0.75   &  10.0       &    60.45733816             &   251.97880982      \\
1.0    &  10.0       &    96.27102702             &   335.74822122    \\
1.25   &  10.0       &   137.37472602             &        \\
1.5    &  10.0       &   181.62662627             &        \\
1.75   &  10.0       &   227.70588636             &        \\
2.0    &  10.0       &   274.39609861             &         \\
2.25   &  10.0       &   320.86226234             &       \\
2.5    &  10.0       &   366.59086983             &        \\
\hline
\end{tabular}
\end{center}
\end{table}

The increase of the polarizability of neon as the inner shell radius is increased can be explored by examining the subshell energies, see table \ref{shellneen}. It is seen that a subshell re-arrangement \cite{rearrange} occurs as the inner shell radius is increased with the 2s subshell becoming more weakly bound than than the 2p subshell as electrons are excluded from penetrating into the region near to the nucleus. As the inner shell radius is increased the binding of the 2s orbital is decreased leading to a large increase in the polarizability.

\begin{table}[ht!]
\small
\begin{center}
\caption{Hartree-fock energies for the 1s, 2s and 2p subshells of neon confined by an impenetrable spherical shell.}
\label{shellneen} 
\vspace{12pt}
\begin{tabular}{lllll}
$r_{\rm in}$ &  $r_{\rm out}$ &   E$_{\rm 1s}$ &  E$_{\rm 2s}$ &  E$_{\rm 2p}$ \\
\hline
0.0  &       10.0     &  -32.77243555  &  -1.93039045   &  -0.85040971    \\
0.05 &      10.0    &   -11.67496109 & -0.59126878 & -1.30650997  \\
0.25 &       10.0    &   -3.72928450  &  -0.24957000   & -1.52410597   \\
0.5   &       10.0    &  -2.18429623 &    -0.18295290 &-1.23595070  \\
0.75 &       10.0    &  -1.56730085  & -0.14556270    &  -1.00631972    \\
1.0   &       10.0    &  -1.22548285 & -0.11740385    &  -0.84030892     \\
\hline
\end{tabular}
\end{center}
\end{table}

\clearpage

\section{Conclusions}
The static dipole polarizabilities of noble gas atoms confined in impenetrable spherical spheres and impenetrable spherical shell potentials has been studied using the RPAE approximation. Confinement  in an impenetrable sphere leads to a decrease in the polarizability as the radius of the sphere is decreased. However confinement by an impenetrable spherical shell potential leads to a large increase in the polarizability as the inner shell radius is increased. This offers the potential of tuning the polarizability of an atomic system by confinement. Topics  in the polarizabilities of confined multi-electron atoms that require further investigation include the influence of relativistic effects \cite{diracfock} and exploration of a range of confining potentials beyond the hard-wall confinement approximation, such as repulsive \cite{swelling1},\cite{swelling2} or attractive \cite{endo1} penetrable spherical shell potentials, with a view to exploring both higher multipole polarizabilities and dynamic polarizabilities. 

\section*{Acknowledgments}
We wish to acknowledge S.~D.~Loch for the generous provision of computing facilities and G.~F.~Gribakin for helpful discussions  and suggestions.\\


\clearpage

\begin{thebibliography}{99}

% confined atom reviews
\bibitem{confinerev1} W.~Jask\'{o}lski {\it Phys. Rep.} {\bf 271}, 1 (1996).
\bibitem{confinerev2} A.~L.~Buchachenko {\it J. Phys. Chem. B} {\bf 105}, 5839 (2001).
\bibitem{confinerev3} J.~P.~Connerade and P.~Kengkan {\it Proc. Idea-Finding Symp. (Frankfurt Institute for Advanced Studies)} p 35 (2003).
\bibitem{confinerev4} V.~K.~Dolmatov, A.~S.~Baltenkov, J.~P.~Connerade and S.~T.~Manson {\it  Radiat. Phys. Chem} {\bf 70}, 417 (2004).
\bibitem{confinerev5}  {\it Theory of Confined Quantum Systems: Part One, Advances in Quantum Chemistry}, edited by J.~R.~Sabin, E.~Br\"{a}ndas and C.~A.~Cruz (Academic Press, New York, 2009), Vol. 57, pp. 1-334.
\bibitem{confinerev6} {\it Theory of Confined Quantum Systems: Part Two, Advances in Quantum Chemistry}, edited by J.~R.~Sabin, E.~Br\"{a}ndas and C.~A.~Cruz (Academic Press, New York, 2009), Vol. 58, pp. 1-297.


% Polarizablity reviews
\bibitem{polarrev1} P. ~Schwerdtfeger {\it Atoms, Molecules and Clusters in Electric Fields: Theoretical Approaches to the calculation of electric polarizability}, edited by G.~Maroulis (World Scientific, London, 2006),  Chap. 1, p. 1.
\bibitem{polarrev2} J.~Mitroy, M.~S.~Safronova and C.~W.~Clark {J. Phys. B: At. Mol. Opt. Phys.} {\bf 43}, 202001 (2010).

% H 
\bibitem{early1} A.~Michels, J.~de~Boer and A.~Bijl {\it Physica} {\bf 4}, 981 (1937).
\bibitem{early2} A.~Sommerfeld and H.~Welker {\it Ann. Phys.} {\bf 424}, 56 (1938).
\bibitem{early3} S.~R.~De~Groot and C.~A.~Ten~Seldam {\it Physica} {\bf 12}, 669 (1946).
\bibitem{h1} P.~W.~Fowler {\it Molec. Phys.} {\bf 53}, 865 (1984).
\bibitem{h2} A.~Banerjee, K.~D.~Sen, J.~Garza, and R.~Vargas {J. Chem. Phys.} {\bf 116}, 4054 (2002).
\bibitem{h3} C.~Laughlin {\it J. Phys. B: At. Mol. Opt. Phys.} {\bf 37}, 4085 (2004).
\bibitem{mont1} H.~E.~Montgomery {\it Chem. Phys. Lett.} {\bf 352}, 529 (2002).
\bibitem{cohen} B.~L.~Burrows and M.~Cohen {Phys. Rev. A} {\bf 72}, 032508 (2005)
\bibitem{mont2} H.~E.~Montgomery and K.~D.~Sen {\it Phys. Lett. A} {\bf 376}, 1992 (2012).
\bibitem{h4} H.~E.~ Montgomery,~Jr. and V.~I.~Pupyshev {\it Eur. Phys. J. H} {\bf 38}, 519 (2013).


% He
\bibitem{he0} C.~A.~Ten~Seldam and S.~R.~De~Groot {\it Physica} {\bf 18}, 905 (1952).
\bibitem{he3} S.~Waugh, A.~Chowdhury and A.~Banerjee {\it J. Phys. B: At. Mol. Opt. Phys.} {\bf 43}, 225002 (2010).
\bibitem{he1} T.~Sako and G.~H.~F.~Diercksen {\it J. Phys. B: At. Mol. Opt. Phys.} {\bf 36 }, 3743 (2003). 
\bibitem{he2} F.~Holka, P.~Neogr\'{a}dy, V.~Kell\"{o} , M.~Urban and G.~H.~F.~Diercksen {\it Molec. Phys.} {\bf 103}, 2747 (2005).
\bibitem{he4} S.~Kar and Y.~K.~Ho {\it Phys. Rev. A} {\bf 80}, 062511 (2009).

% H cylinder
\bibitem{hcylinder} S.~A.~Ndengu\'{e}, O.~Motapon, R.~L.~Melingui Melono and A.~J.~Etindele {\it J. Phys. B: At. Mol. Opt. Phys.} {\bf 47 }, 015002 (2014).

% He spheroid
\bibitem{hespheroid} A.~Corella-Madue\~{n}o, R.~A.~Rosas, J.~L.~Mar\'{i}n, R.~Riera {\it Int. J. Quant. Chem.} {\bf 77}, 509 (2000).

% Alkali Metals 
\bibitem{alkali1} S.~H.~Patil, K.~D.~Sen, and Y.~P.~Varshni {\it Can. J. Phys.} {\bf 83}, 919 (2005).
\bibitem{db1} N.~Li-Na and Q.~Yue-Ying {\it Chinese Phys. B} {\bf 21}, 123201 (2012).
\bibitem{db2} H.~W.~Li and S.~Kar {\it Phys. Plasmas} {\bf 19}, 073303 (2012).
\bibitem{db3} Yue-Ying~Qi and Li-Na~Ning {\it Phys. Plasmas} {\bf 21}, 033301 (2014). 


% Be
\bibitem{be} K.~Strasburger and P.~Naci\c{a}\.{z}ek {\it J. Phys. B: At. Mol. Opt. Phys.} {\bf 47}, 025002 (2014).

\bibitem{ar0}  C.~A.~Ten~Seldam and S.~R.~De~Groot {\it Physica} {\bf 18}, 910 (1952).
\bibitem{hartree3} P.~K.~Chattaraj and U.~Sarkar {\it J. Phys. Chem. A} {\bf 107}, 4877 (2003).
\bibitem{hartree4} P.~K.~Chattaraj and U.~Sarkar {\it Chem. Phys. Lett.} {\bf 372}, 805 (2003).
\bibitem{hartree5} P.~K.~Chattaraj, B.~Maiti and U.~Sarkar {\it Proc. Indian Acad. Sci. (Chem. Sci.)} {\bf 115}, 195 (2003).
\bibitem{hartree6} M.~van~Faassen {\it J. Chem. Phys.} {\bf 131}, 104108 (2009).

% Noble gas experiments
\bibitem{nobleexp} M.~Lallemand and D.~Vidal {\it J. Chem. Phys.}  {\bf 66}, 4776 (1977).

% endofullerenes
\bibitem{endo1} S.~A.~Ndeng\'{e} and O.~Motapon {\it J. Phys. B: At. Mol. Opt. Phys.} {\bf 41}, 045001 (2008).
\bibitem{endo2} D.~Sh.~Sabirov and R.~G.~Bulgakov {\it JETP Letters} {\bf 92}, 662 (2010).
\bibitem{endo3} H.~Yan, S.~Yu, X.~Wang, Y.~He, W.~Huang, M.~Yang {\it Chem. Phys. Lett.} {\bf 456}, 223 (2008).

\bibitem{amusia73} M.~Ya.~Amusia, N.~A.~Cherepkov and S.~G.~Shapiro {\it Soviet Physics JETP} {\bf 36}, 468 (1973).
\bibitem{amusia75} M.~Ya.~Amusia and N.~A.~Cherepkov, {\it Case Studies in Atomic Physics} {\bf 5}, 47 (1975).
\bibitem{amusia90} M.~Ya.~Amusia, {\it Atomic Photoeffect} (Plenum Press, 1990). 
\bibitem{noblecc} T.~Nakajima and K.~Hirao {\it Chem. Lett.} {\bf 30}, 766 (2001).
\bibitem{rrpa} W.~R.~Johnson, D.~Kolb and K.~Huang {\it At. Data Nucl. Data Tables} {\bf 28}, 333 (1983).

% Numerical methods
\bibitem{hfcode1} M~ Ya.~Amusia and L~ V.~Chernysheva, {\it Automated system for atomic structure calculations} (Nauka, Leningrad, 1983).
\bibitem{hfcode2} L~ V.~Chernysheva, N.~A.~Cherepkov  and V.~Radojevi\'{c} {\it Computer Phys. Commun.} {\bf 11}, 57 (1976).
\bibitem{bspline} H.~Bachau , E.~Cormier, P.~Decleva, J.~E.~Hansen and F.~Mart\'{i}n {\it Rep. Prog. Phys.} {\bf 64}, 1815 (2001).
\bibitem{deboor} C.~De.~Boor {\it A Practical guide to Splines} ((Springer, New York, 1978),
\bibitem{ludlowphd} J.~Ludlow. {\it Many-body theory of Positron-Atom interactions}. PhD thesis, Queen's University Belfast, Department of Mathematics and Theoretical Physics (2003).

% Mainly analytical in nature/early work
\bibitem{early4} R.~A.~Buckingham {\it Proc. Roy.Soc. A} {\bf 160}, 94 (1937).
\bibitem{early5} J.~G.~Kirkwood {\it Phys. Z.} {\bf 33}, 57 (1932).

\bibitem{hartree1} E.~V.~Lude\~{n}a {\it J. Chem. Phys.} {\bf 69}, 1770 (1978).
\bibitem{hylleraas} G.~\L ach, B.~Jeziorski and K.~Szalewicz  {\it Phys. Rev. Lett.} {\bf 92}, 233001 (2004).

\bibitem{g3data} g3data program (http://www.frantz.fi/software/g3data.php).

\bibitem{chbucky} J.~A.~Ludlow, Teck-Ghee~Lee and M.~S.~Pindzola, {\it J. Phys. B: At. Mol. Opt. Phys.} {\bf 43}, 235202  (2010).

% h shell confined
\bibitem{hshell1} K.~D.~Sen, J.~Garza, R.~Vargas and N.~Aquino {\it Physics Letters A} {\bf 295}, 299 (2002).
\bibitem{hshell2} K.~D.~Sen {\it J. Chem. Phys.} {\bf 122}, 194324 (2005).
\bibitem{hshell3} B.~L.~Burrows and M.~Cohen {\it Molec. Phys.} {\bf 106}, 267 (2008).
\bibitem{hshell4} N.~Aquino {\it Molec. Phys.} {\bf 107}, 2319 (2009).
\bibitem{hshell5} B.~L.~Burrows and M.~Cohen {\it Molec. Phys.} {\bf 107}, 2323 (2009).

% Subshell filling
\bibitem{rearrange} J.~P.~Connerade, V.~K.~Dolmatov and P.~A.~Lakshmi {\it  J. Phys. B: At. Mol. Opt. Phys.} {\bf 33}, 251 (2000).

% Relativistic
\bibitem{diracfock} J.~P.~Connerade and R.~Semaoune {\it  J. Phys. B: At. Mol. Opt. Phys.} {\bf 33}, 3467 (2000).

% atomic swelling
\bibitem{swelling1} V.~K.~Dolmatov and J.~ L.~King {\it J. Phys. B: At. Mol. Opt. Phys.} {\bf 45}, 225003 (2012).
\bibitem{swelling2} V.~K.~Dolmatov {\it J. Phys. B: At. Mol. Opt. Phys.} {\bf 46}, 095005 (2013).

\end{thebibliography}
\end{document}